\documentstyle[aps,multicol,epsf]{revtex}
\begin{document}
\draft
\title{Numerical test of the damping time of layer-by-layer growth \\ 
on stochastic models \\} 
\vspace{1cm}
\author{S. Park$^{1}$, H.~Jeong$^{2}$ and B.~Kahng$^{1}$ \\} 
\address{
$^1$ Department of Physics and Center for Advanced Materials and 
Devices, \\
Kon-Kuk University, Seoul 143-701, Korea \\
$^2$ Department of Physics, University of 
Notre Dame, Notre Dame, IN 46556, U.S.A. \\}
\maketitle 
\thispagestyle{empty}

\begin{abstract} 
We perform Monte Carlo simulations on stochastic models 
such as the Wolf-Villain (WV) model and the Family model 
in a modified version to measure mean separation $\ell$ 
between islands in submonolayer regime and damping time 
$\tilde t$ of layer-by-layer growth oscillations on one dimension. 
The stochastic models are modified, allowing diffusion within 
interval $r$ upon deposited. 
It is found numerically that the mean separation 
and the damping time depend on the diffusion interval $r$, 
leading to that the damping time is related to the mean 
separation as ${\tilde t} \sim \ell^{4/3}$ for the WV model and 
${\tilde t} \sim \ell^2$ for the Family model. 
The numerical results are in excellent agreement with recent 
theoretical predictions.    
\end{abstract} 

\begin{multicols}{2}
\narrowtext
Recently problems of surface growth by molecular beam epitaxy (MBE) 
have been attractive in statistical physics [1]. 
From the point of view of statistical physics, 
it is intriguing how various stochastic processes involving many 
atoms such as shot noise, diffusion, and nucleation, 
are described in terms of scaling structures. These cooperative phenomena 
depend crucially on the ratio between the diffusion constant $D$ 
and the deposition rate $F$. The deposition rate 
is defined as number of atoms landing on surface 
per unit area and unit time. 
Atoms deposited on surface diffuse until they meet one another 
to form dimers which then grow into islands of monoatomic
height. The mean distance of nucleation events, corresponding 
to the mean separation between islands, is determined by the ratio  
between the diffusion constant $D$ and the deposition rate $F$ as [2-8]  
\begin{equation}
\ell \sim (D/F)^{\gamma}. 
\end{equation}
The exponent $\gamma$ depends on diffusion process of adatoms and 
islands shape. It is also a function of the critical island size 
$i^*$, defined by the size $i^*+1$ of the smallest 
island which is stable enough that it never decays before 
capturing next adatom. When only adatoms can move and desorption 
can be neglected, the exponent $\gamma$ can be predicted using the 
kinetic theory as [6] 
\begin{equation}  
\gamma = {i^* \over {2 i^* + d + d_f}}. 
\end{equation} 
However, recently the formula was corrected for the 
case of $d=1$ and $i^* \ge 2$ as [9] 
\begin{equation}  
\gamma = {i^* \over {2 i^* + 3}}. 
\end{equation} 

As number of adatoms increases in submonolayer, islands grow in their sizes 
and coalesce into forming bigger islands, and eventually 
cover a monolayer. If adatoms are allowed to diffuse to stable sites 
such as kink sites or step edges before other adatoms getting deposited 
and if interlayer diffusion is not inhibited by the Ehrlich-Schwoebel 
barrier [10,11], smooth surface with a minimum of defects is grown. 
In this case, surface exhibits a layer-by-layer growth [12]. 
As islands evolve, the density of atomic steps oscillates, 
which is a hallmark of layer-by-layer growth. 
The oscillating behavior can be monitored through the 
RHEED intensity in laboratory, which is useful for 
being aware of film thickness easily. 
Layer-by-layer growth is important for the fabrication 
of microelectronic devices. 
It is known that the oscillation persists perfectly 
when system size is smaller than the layer-coherence length 
$\tilde \ell$ such that for two sites within the coherence 
length $\tilde \ell$, they grows coherently, and beyond 
which they are out of phase [13]. 
The coherence length is related to the separation between 
islands as  
${\tilde \ell} \sim {\ell^{4/(4-d)}}$. 
The coherence length is much larger than 
other length scales such as the islands separation $\ell$ or 
the characteristic length $\ell_0 \sim (D/F)^{1/(2+d)}$, 
where $\ell_0$ is formed by a dimensional combination of $D$ 
and $F$, meaning that if system size is smaller than $\ell_0$, 
it can accommodate at most one island.  
When system size is larger than $\tilde \ell$, the 
oscillation of step density is damped due to various types of 
fluctuations. 
It is recently found [13] that there exists a characteristic 
time $\tilde t$, beyond which the damped oscillation disappears. 
The characteristic time is scaled as 
\begin{equation}
{\tilde t} \sim (D/F)^{\delta},
\end{equation}           
meaning the critical time for the transition from layer-by-layer 
growth to kinetic roughening growth. \\ 

In the kinetic roughening growth, when number of adatoms 
is conserved, surface growth may be described 
by the equation, [11,13,14,15]   
\begin{equation}
\partial_t h=- \nabla \cdot{}{\bf j} + \eta ,
\end{equation}
where $h$ means surface height, $\bf j$ is adatom current, 
and $\eta({\bf x},t)$ denotes shot noise satisfying 
$\langle \eta({\bf x},t) \rangle =0$ and having the correlation, 
\begin{equation}
\langle \eta({\bf x},t) \eta({\bf x'},t') \rangle \sim  
\delta^d({\bf x}-{\bf x'}) \delta(t-t') , 
\end{equation}
with substrate dimension $d$.
When adatom current is driven by the gradient of surface curvature 
and square of surface tilt with coefficients $K$ and $\lambda$ 
[11,13,14,15], 
\begin{equation}
{\bf j} = \nabla [K \nabla^2 h - \lambda (\nabla h)^2]. 
\end{equation}
The continuum equation for surface growth is written as  
\begin{equation}
\partial_t h = -K \nabla^4 h +\lambda \nabla^2 (\nabla h)^2 + 
\eta(x,t), 
\end{equation}
which, called the conserved Kardar-Parisi-Zhang (cKPZ) equation, 
has been discussed in the context of MBE growth.  
For the cKPZ universality class, it was derived that the exponent 
$\delta$ for the damping time is related to $\gamma$ for the 
islands-separation as [13] 
\begin{equation}
{\delta \over \gamma} = {{4d} \over {4-d}}. 
\end{equation}
The derivation was based on dimensional analysis on the continuum 
equation. The formula was checked numerically in one dimension 
performing the coarse graining Monte Carlo simulations [13]. \\ 

For the surface growth driven by downhill current with coefficient 
$\nu$,  
\begin{equation}
{\bf j} = -\nu \nabla h , 
\end{equation}
the continuum equation is written as 
\begin{equation}
\partial_t h = \nu \nabla^2 h + \eta(x,t), 
\end{equation}
which is called the Edwards-Wilkinson (EW) equation [16]. 
In this case, adatoms tend to move in downward direction on 
surface, which is realized by exchanging mobile atom 
with adatom at step edge as observed in homo epitaxial 
growth on Ir(111) [17].    
In this case, the exponent $\delta$ for the damping time 
is related to $\gamma$ as [13]
\begin{equation}
{\delta \over \gamma} = {{2d} \over {2-d}}, 
\end{equation}
which was also derived based on dimensional analysis 
on the continuum equation, Eq.(11). However, successful numerical 
confirmation of Eq.(12) has not been reported yet. 
Our preliminary numerical result for MBE growth with downhill 
current does not fit well to Eq.(12), requesting numerical check 
of the formulae, Eqs.(9,12) on stochastic models 
belonging to the cKPZ and EW universalities, respectively.  
In this letter, we perform Monte Carlo simulations on the 
stochastic models, the Wolf-Villain (WV) model [18] 
and the Family model [19] 
with some modifications in their dynamic rules to check Eqs.(9,12).\\
 
Let us first consider the modified WV model, of which 
the dynamic rule is defined as follows.  
First, a site, say $i$-th site, is selected randomly on a 
one-dimensional flat substrate with system size $L$. 
Then we consider a subset of the system, 
composed of $2r+1$ sites, the randomly selected site, 
and its $2r$ neighboring sites on its right and left sides,
respectively, within distance $r$. 
Among the $2r+1$ sites, surface is 
advanced at the site offering largest binding, 
that is, most occupied neighbors. If sites 
offering largest binding exists more than one, 
then a site closest to the $i$-th site among them is taken 
and its height is increased by one. The case of $r=1$ is 
reduced to the original version of the WV model. 
If the distance $r$ is regarded as diffusion length, 
then the diffusion constant would be related to the distance 
as $D \sim r^2$.  
We measure the density of islands in submonolayer regime 
and the surface fluctuation width $W^2$ [20], 
\begin{equation}
W^2(L,t)= \langle {1\over L} \sum_{i}(h_i-\bar{h})^2
\rangle,
\end{equation} 
with $\bar{h}=\sum_i h_i /L$ as varying the diffusion 
length $r$. Note that the density of islands $\rho$ is related 
to the mean separation $\ell$ as $\rho \sim \ell^{-1}$ in one 
dimension, and the damped oscillation in the step density 
occurs in the surface fluctuation width in the same way. 
In our simulations, we used the noise reduction method 
with the noise reduction parameter $m=5$ to get better data. 
However, we did not vary the parameter $m$ in any case, 
so that the noise reduction parameter does not play a role 
of control parameter as considered in Refs. 21-22.  
As shown in Fig.1, the density of islands well behaves as 
$\rho \sim r^{-0.94}$ for large values of $r$, suggesting that 
the exponent $\gamma \approx 0.47$. The numerical result is 
close to the theoretical result $\rho \sim r^{-1}$, 
which is obtained as follows.   
If we divide the system into small cells 
with equal size $2r+1$, and a particle is deposited 
on a cell randomly, then one island would be formed 
on each cell, and the separation of islands in the system 
varies as $\ell \sim r$ for large $r$. 
Thus $\rho \sim r^{-1}$. However, 
the cell boundary is virtual and could be overlapped to 
each other in our simulations, but the correction due to the 
fluctuation of boundaries would be of higher 
order. Thus the density of islands would behave as 
$\rho \sim r^{-1}$ up to leading order, and 
$\gamma \approx 1/2$. We also count number of 
nucleation event by ignoring monomers, and the result 
obtained is the same as Fig.1. 
Next, the surface fluctuation width $W^2$ exhibits a damped 
oscillation as shown in Fig.2(a). 
The data for different diffusion lengths are well collapsed, 
even for small values of $r$,  
when time is rescaled as $t/{\tilde t}$ with 
${\tilde t} \sim r^{4/3}$ as shown in Fig.2(b). 
The rescaled time suggests $\delta = 2/3$. 
Therefore, the ratio of the two exponents $\gamma$ 
and $\delta$ is obtained as $\delta/\gamma \approx 4/3$, which is 
consistent with the theoretical prediction, Eq.(9) for $d=1$. \\ 

Next, we consider the modified Family model. In this case, 
we choose a site randomly, say $i$-th site, on one dimension 
substrate with system size $L$, and consider a subset 
of $2r+1$ sites as before. 
Among the $2r+1$ sites, surface is advanced at lowest site. 
If sites with the lowest height exist more than one, 
a site closest to the $i$-th site is taken. 
The case of $r=1$ is reduced to the original version of the 
Family model. We vary the distance $r$, and examine the 
density of islands and the damped oscillating behavior 
of the surface fluctuation width. The density of 
islands in submonolayer regime behaves as the case of 
the WV model. For the surface fluctuation width,   
the data are well collapsed even for small values of 
$r$ when time is rescaled as $t/r^2$ as shown in Fig.3(a), 
implying that the damping time behaves as ${\tilde t} \sim r^2$. 
Thus the ratio between the exponents $\delta$ and 
$\gamma$ is obtained as $\delta/\gamma \approx 2$ 
for the Family model, which is consistent with the theoretical 
prediction, Eq.(12). 
We also check the case involving a hump on the way to the lowest 
site in the subset. In Fig.3(a), we considered the case that 
adatom can run over the hump, and stay on the lowest site 
within the interval. 
In Fig.3(b), adatom cannot run over hump, and is allowed 
to move only in a descending way or on the same height. 
We cannot find any difference between the two in the collapsing 
behavior, implying the assumption of the dephasing 
length $\tilde \ell$, within which the layer difference is 
at most one, is valid. Therefore, we conclude that the 
theoretical prediction is correct for the stochastic models, 
the WV model and the Family model, belonging to the 
cKPZ and EW universalities.  
However, the relevance of the theoretical prediction 
to the MBE growth with downhill current 
is to be further investigated.\\

In summary, we have performed Monte Carlo simulations 
on the Wolf-Villain model and the Family model involving 
the diffusion length $r$ on one dimension. 
We measure the density of islands $\rho$ in submonolayer 
regime and the damping time $\tilde t$ for the oscillation of 
layer-by-layer growth as varying the diffusion length. 
We obtained numerically that the exponents $\delta$ 
and $\gamma$ describing the damping time and the density 
of islands are related to each other as $\delta/\gamma \approx 4/3$ 
for the Wolf-Villain model and $\delta/\gamma \approx 2$ for the Family model. 
This result is in excellent agreement with recent 
theoretical predictions. \\  

One of authors (B.K.) would like to thank D.E. Wolf for 
helpful discussions. This work is supported in part 
by the KOSEF (971-0207-025-2), and through the SRC 
program of SNU-CTP.

\begin{figure}
\centerline{\epsfxsize=8.3cm \epsfbox{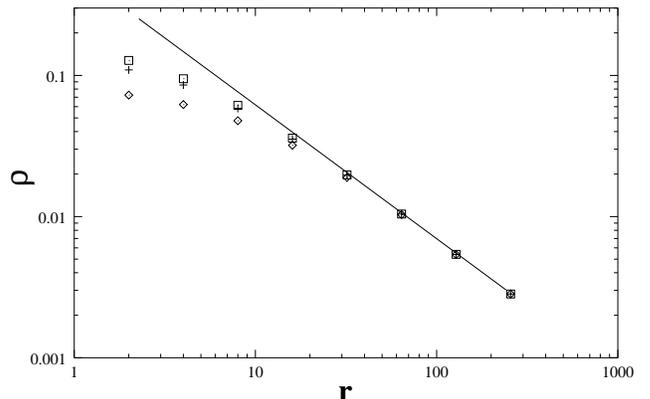}}
\caption{Double logarithmic plot of density of islands $\rho$ 
versus diffusion length $r$ for the Wolf-Villain model 
in one dimension. 
The data are obtained from system size $L=1000$ at different 
coverages, $0.1$, $0.2$, and $0.3$ from the bottom, 
and are averaged over 1000 runs. 
The solid line with slope $-0.94$ is the guidance to the eyes.} 
\label{fig1}
\end{figure}

\begin{figure}
\centerline{\epsfxsize=8.3cm \epsfbox{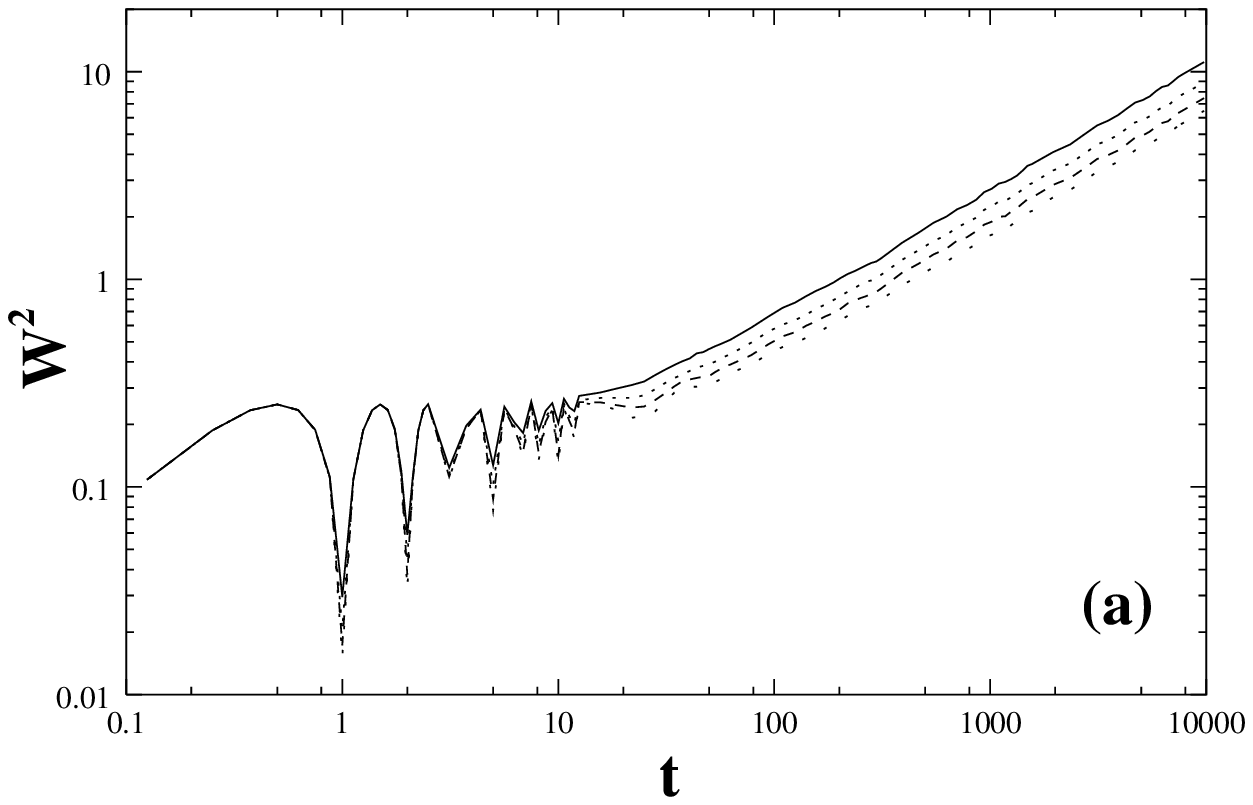}}
\centerline{\epsfxsize=8.3cm \epsfbox{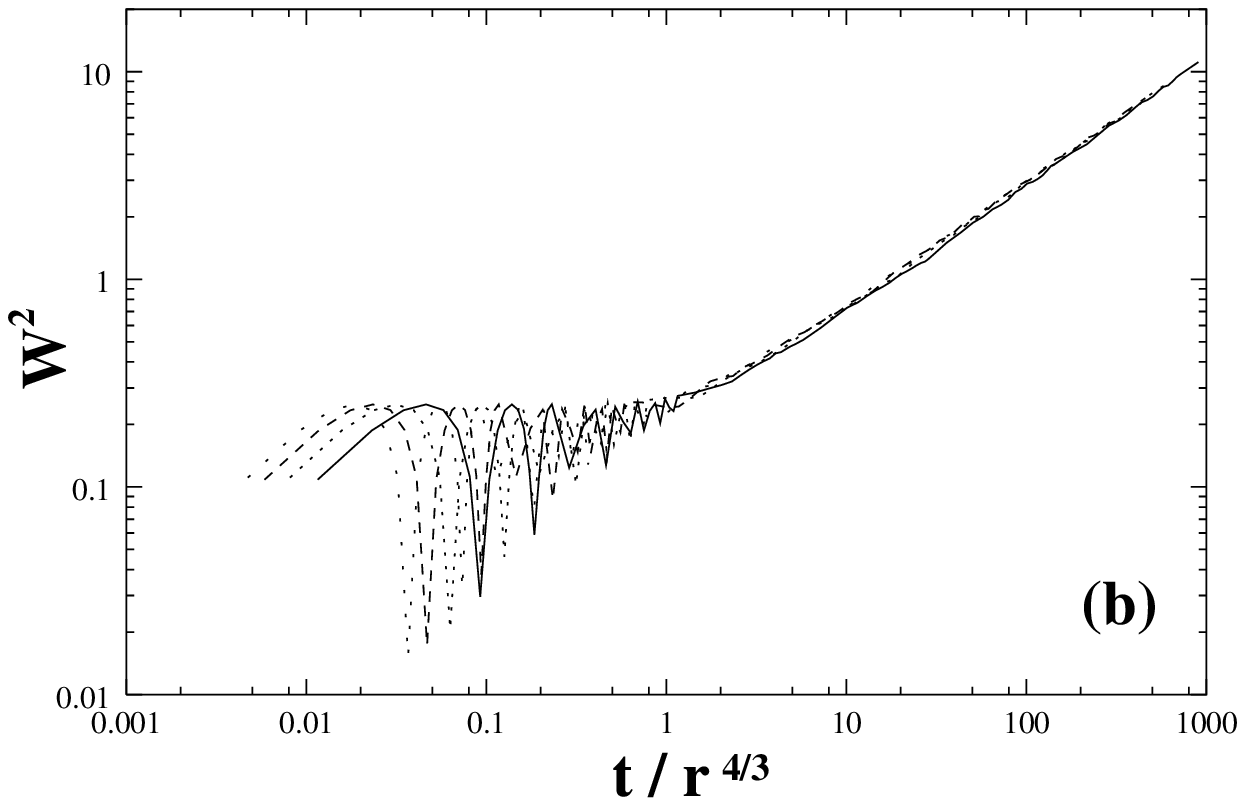}}
\caption{Double logarithmic plot of the surface 
fluctuation width versus time $t$ for the case 
(a) and rescaled time $t/r^{4/3}$ for the 
case (b) in the Wolf-Villain model. 
The data are obtained from system size $L=1000$, and 
are averaged over 100 runs. 
The diffusion lengths $r=6,8,10$ and 12 were used 
from the top.}
\label{fig2}
\end{figure}

\begin{figure}
\centerline{\epsfxsize=8.3cm \epsfbox{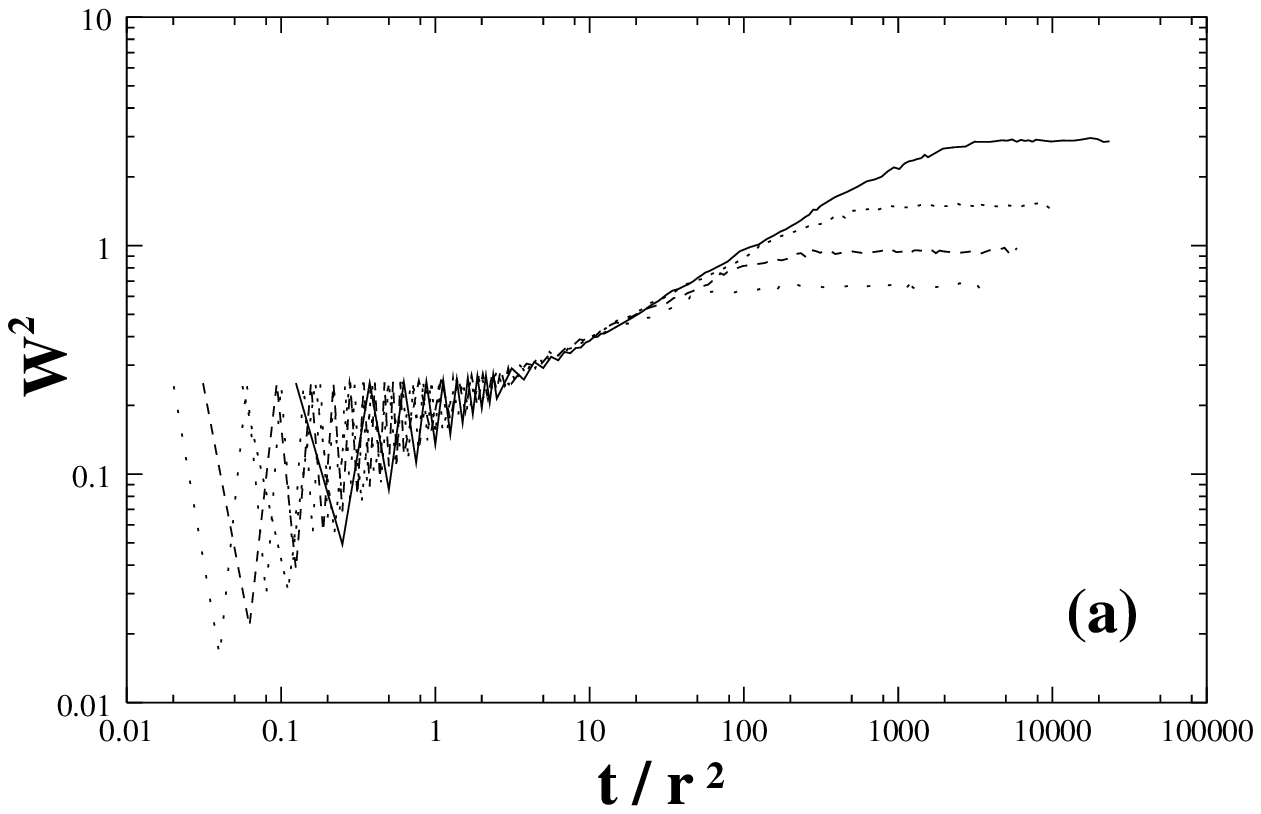}}
\centerline{\epsfxsize=8.3cm \epsfbox{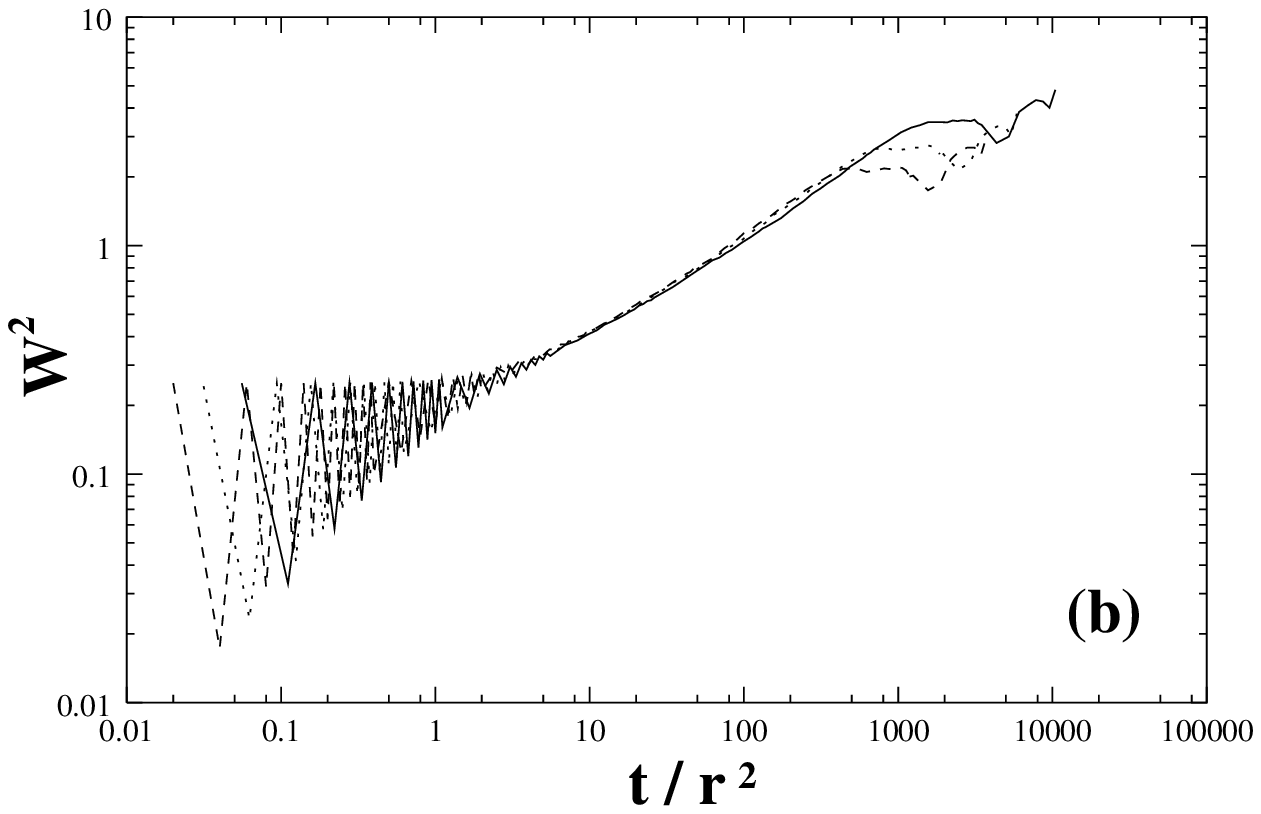}}
\caption{Double logarithmic plot of the surface fluctuation 
width $W^2$ versus rescaled time $t/r^{2}$ in the Family model. 
In simulations, adatoms are allowed to run over humps 
for the case (a) and are not allowed to do for the case (b).
For the case (a), various diffusion lengths, 
$r=2,3,4$ and 5 are used, and for the case 
(b), $r=3,4$ and $5$ are used. The data are obtained 
from system size $L=1000$, and are averaged over 100 runs.} 
\label{fig3}
\end{figure}

\end{multicols}
\end{document}